# EQUILIBRIUM FLUCTUATIONS IN A METASTABLE STATE OF A GINZBURG-LANDAU SYSTEM

D. I. Uzunov

Collective Phenomena Laboratory, G. Nadjakov Institute of Solid State Physics, Bulgarian Academy of Sciences, BG-1784 Sofia, Bulgaria

and A. Umantsev

Department of Chemistry/Physics, Fayetteville State University, Fayetteville, NC 28301

December 3, 2015

ABSTRACT

We calculate thermal fluctuation properties—volume-averaged order parameter, Helmholtz free and internal energies, and their variances—of a supersaturated disordered phase in the Gibbs canonical ensemble for an asymmetric (third-order interactions), athermal (independence of the supersaturation and thermal noise) effective Hamiltonian. These properties are different from those of the symmetric thermal one with the most important differences being the phase coexistence and 'thermal expansion'. The fluctuation properties of the system were calculated theoretically, using the perturbation method, and numerically, using the 'brute force' simulations method. Overall, the numerical calculations match the theory within the accuracy of the numerical method. However, a discrepancy of the dependence of the internal energy and its variance on the supersaturation exists. Results of the present study can be used for calculations of the fluctuation properties of the systems and modeling of nucleation and other rare events in the framework of the Ginzburg-Landau method.

I. INTRODUCTION

Fluctuations in metastable states and ensuing phase transitions are important in many areas of physics. Quantum fluctuations due to Heisenberg's uncertainty principle could have produced small perturbations in the space-time that shaped our Universe. Phase transitions occurred in the early Universe as it cooled could have produced the forces and particles we observe today. On the other hand, *thermal fluctuations* (fluctuations in a system subjected to thermal noise) are important in many practical situations. Although metastable states of matter e.g., supercooled or superheated water, diamond, etc, survive for a period of time, eventually they transfer to equilibrium states: ice, steam, graphite, etc. For the period of time while the metastable states survive the decomposition, they may be considered as equilibrium ones and their fluctuations bear all the features of equilibrium fluctuations.

From the stand point of the theory of phase transitions the abovementioned examples belong to the category of first-order transitions, which take place away from the point of equilibrium. A theoretically coherent approach to consider these processes is the framework of a *Ginzburg-Landau* (GL) model [1-6]. It is based on the concepts of an *order parameter* (OP) as the main characteristic of the state and *effective Hamiltonian* as the main characteristic of the interactions. The GL method allows one to treat the fluctuations using two different approaches—equilibrium and dynamic—whose consistency has been verified many times [1-4, 6]. According to Penrose and Lebowitz [7], a *metastable state* is a state, in the vicinity of which the representative point of the system spends long time before eventually leaving it with low probability of return. To ensure metastability of a state in the former approach we apply only small fluctuations, which allows us to use the perturbation theory. In the latter approach to preserve the metastability, we restrict the fluctuation time by the lifetime of the state.

Traditionally, thermal fluctuations have been studied in the framework of a symmetric Hamiltonian, which has the OP monomials of the even powers (e.g. second and fourth) in the truncated power-series expansion [1-5, 8]. Here the "Gaussian" (quadratic, harmonic) term describes free (non-interacting) fluctuations whereas the terms of higher order (quartic, anharmonic) describe the relevant fluctuation interactions in the system. However, this type of Hamiltonian does not include the *non-critical* phenomena of phase coexistence, nucleation, growth and coarsening. A standard trick to study these phenomena in the GL framework is to apply an external field, e.g. electric, magnetic, etc., which breaks the symmetry of the OP space in favor of one orientation. Unfortunately, many material systems and transformations, e.g. crystallization, do not have physical quantities which play the role of the external field. In this case, a standard way to study the phenomena of phase coexistence, nucleation, growth, and coarsening in the GL framework is to consider an *asymmetric* Hamiltonian, which contains the third-power monomial in the power-series expansion [6, 9-14]. Although having many similar properties with the symmetric one, the fluctuation properties of the asymmetric Hamiltonian are different because its leading anharmonic (non-Gaussian) term has lower order than that of the symmetric one.

Another important feature of the phase transformations in materials is that very often they are driven not only by temperature changes but also by changes of pressure, concentration, etc. This feature is modeled here by *athermal* Hamiltonian that is, the Hamiltonian with temperature independent (or non-critically dependent) coefficients of the singular part. This also changes the fluctuation properties of the system subjected to thermal noise.

An effective tool of resolving complex dynamical problems where the spatially distributed fluctuations are important is the method of numerical simulations [15, 16]. In this publication we calculate the fluctuation properties—order parameter, internal and free energies, etc.—of equilibrium (stable or metastable) states of a system described by the asymmetric, athermal Hamiltonian in the Gibbs canonical ensemble using both methods: the equilibrium perturbation theory (Sec. II) and numerical simulation of the dynamical system (Sec. III). Then we compare the results and draw the conclusions (Sec. IV).

## II. EQUILIBRIUM PERTURBATION THEORY

### 1. Ginzburg-Landau Hamiltonian and Phase Diagram

We consider a typical first order phase transition, e.g., crystallization or order-disorder, caused not by temperature changes but by changes of pressure, concentration, or another thermodynamic quantity. These changes will be described by a dimensionless *supersaturation* $\Delta$, which is a quantitative expression of the system's deviation from the equilibrium. The states of the system will be described by the scalar order parameter field $\eta(\mathbf{x})$, $\mathbf{x}\in V$ where V is volume of the system considered to be constant. Fluctuations of the OP field can be described by the GL effective Hamiltonian [1-6, 8]:

$$\mathcal{H}\{\eta\} = \int_V d^3x \left\{ H(\eta; \Delta) + \frac{1}{2}\kappa(\boldsymbol{\nabla}\eta)^2 \right\} \tag{1}$$

which consists of the uniform contribution described by the density $H(\eta; \Delta)$ and the contribution of the spatial fluctuations proportional to the stiffness parameter $\kappa$, which depends on the radius of interactions responsible for the phase transition. In this publication we use the following expression of the effective Hamiltonian density [6, 10-12]:

$$H(\eta; \Delta) = \frac{1}{2}W \left\{ (1-\Delta)\eta^2 - 2\left(1 - \frac{\Delta}{3}\right)\eta^3 + \eta^4 \right\} \tag{2}$$

where W is the energy density scale, see Figure 1. As known [1], the Helmholtz free energy of a uniform system described by the effective Hamiltonian, Eqs.(1, 2), is expressed as follows:

$$F(\eta; \Delta) = F_\alpha + H(\eta; \Delta)V \tag{3}$$

The uniform system described by Eqs.(2, 3) has the following equilibrium ($\partial F(\eta_e; \Delta)/\partial\eta=0$) states $\eta_e$: 'disordered' $\alpha$-phase with $\eta_\alpha=0$, 'ordered' $\beta$-phase with $\eta_\beta=1$, and the transition state between the $\alpha$ and β phases with $\eta_t=(1-\Delta)/2$. Coefficients of the effective Hamiltonian density, Eq.(2), guarantee that the order parameters of the α and β phases are independent of the supersaturation $\Delta$ (in Ref. 6 such density was called *tangential*). The non-fluctuating free energy of the α-phase $F_\alpha$ is a function of temperature, the free energy of the β-phase is $F_\beta = F_\alpha - VW\Delta/6$, and the free energy of the transition state is $F_t = F_\alpha + VW(3+\Delta)(1-\Delta)^3/96$. For the asymmetric density, Eq.(2), $\Delta=0$ is the phase equilibrium point, $\Delta=1$ is the α-phase *spinodal* point, and for $0<\Delta<1$ the α-phase is metastable, see Figure 1. In the vicinity of the spinodal point that is, at ($\eta_\alpha=0$, $\Delta\to 1$) the fluctuation properties of our system are reminiscent of those of the system described by the symmetric Hamiltonian near the critical point; for instance, the fluctuations of the disordered α-phase are large due to flatness of the Hamiltonian density (see Fig. 1). However, in this work we are not specifically concerned with the region near the spinodal point because this region is not easily accessible experimentally and the numerical simulations of Sec. III are not 'tuned' to this region.

2. Fluctuation Theory

Presence of the thermal noise in the system leads to spatial variations of the order parameter $\eta(\mathbf{x}) = \eta_e + \delta\eta(\mathbf{x})$ where $\delta\eta(\mathbf{x})$ is the fluctuation. To analyze the fluctuation effects on the disordered α-phase we set in Eqs.(1-3) $\eta_e = \eta_\alpha = 0$, $\eta=\delta\eta(\mathbf{x})$, and consider the effective Hamiltonian, which, for the purposes of the perturbation theory, is divided into the following contributions:

$$\mathcal{H}\{\eta\} = \mathcal{H}_0\{\eta\} + \mathcal{H}_{int}\{\eta\} \tag{4}$$

Here the Gaussian contribution is

$$\mathcal{H}_0\{\eta\} = \frac{1}{2}\int_V d^3x \left[W(1-\Delta)\eta^2 + \kappa(\nabla\eta)^2\right] \tag{4a}$$

and the fluctuation-interaction contribution is $\mathcal{H}_{int} \equiv \mathcal{H}_3 + \mathcal{H}_4$ where

$$\mathcal{H}_3\{\eta\} = \frac{\Delta-3}{3}W \int_V d^3x\, \eta^3, \tag{4b}$$

$$\mathcal{H}_4\{\eta\} = \frac{1}{2}W \int_V d^3x\, \eta^4. \tag{4d}$$

In the Gibbs canonical ensemble, the statistical average of a thermodynamic quantity $Q\{\eta\}$ is expressed as follows:

$$\langle Q \rangle \equiv \frac{1}{\mathcal{Z}} \int D\eta \, Q\{\eta\} \, e^{-\beta \mathcal{H}\{\eta\}}, \tag{5}$$

where $\beta = 1/k_B T$, $k_B$ is Boltzmann's constant, T is the temperature of the ensemble,

$$\mathcal{Z} \equiv \int D\eta \, e^{-\beta \mathcal{H}\{\eta\}}, \tag{6}$$

is the fluctuation partition function of the system, and $\int D\eta \equiv \prod_{\mathbf{x} \in V} \int D\eta(\mathbf{x})$ denotes the functional integration over all possible configurations of the OP field. Below we calculate the ensemble averages of the volume-averaged OP field:

$$\eta_V \equiv \frac{1}{V} \int_V d^3 x \, \eta(\mathbf{x}), \tag{7}$$

which is proportional to the total order of the system, its variance:

$$Var(\eta_V) \equiv \langle (\eta_V - \langle \eta_V \rangle)^2 \rangle = \langle \eta_V^2 \rangle - \langle \eta_V \rangle^2, \tag{8}$$

and the Helmholtz free energy F=$F_\alpha$+$\mathcal{F}$ where:

$$\mathcal{F} \equiv -\beta^{-1} \ln \mathcal{Z}. \tag{9}$$

is its fluctuation part. Here and below the cursive letters are used for the fluctuation properties.

As known [1, 17], the ensemble averages of the internal energy of fluctuations $\mathcal{E}$ and its variance Var($\mathcal{E}$) (which is related to the specific heat of fluctuations) can be calculated from the Helmholtz free energy, Eq.(9), using the thermodynamic differentiation. That is, the total energy is E=($F_\alpha$−T$\partial F_\alpha$/$\partial$T)+$\mathcal{E}$ where:

$$\mathcal{E} \equiv \langle \mathcal{H} \rangle = \frac{\partial (\beta \mathcal{F})}{\partial \beta} \tag{10}$$

$$Var(\mathcal{E}) \equiv \langle (\mathcal{H} - \langle \mathcal{H} \rangle)^2 \rangle = -\frac{\partial^2 (\beta \mathcal{F})}{\partial \beta^2} \tag{11}$$

The thermodynamic differentiation cannot be used for the volume-averaged OP of the asymmetric Hamiltonian or its variance because the external field in such system, Eqs.(1, 2), is not defined.

For the effective Hamiltonian, Eqs.(1, 2, 4), the functional integrals in Eqs.(5, 6) cannot be calculated exactly. That is why we use the standard perturbation theory of interacting fluctuations [2-5]. That is, first, we consider the exactly soluble Gaussian approximation of non-interacting OP fluctuations ($\mathcal{H}_{int} = 0$). Next, we apply the perturbation-theory expansion, which accounts for the fluctuation interactions:

$$\langle Q \rangle = \frac{\langle Q\, e^{-\beta \mathcal{H}_{int}\{\eta\}} \rangle_0}{\langle e^{-\beta \mathcal{H}_{int}\{\eta\}} \rangle_0} = \langle\langle Q \rangle\rangle - \beta \langle\langle Q \mathcal{H}_{int} \rangle\rangle + \frac{\beta^2}{2} \langle\langle Q\, \mathcal{H}_{int}^2 \rangle\rangle + \cdots \quad (12)$$

Then we truncate the series on the term of second order in $\beta\mathcal{H}_{int}$ and neglect the terms of higher orders. Here we used the following expressions for the connected «» and regular Gaussian <>$_0$ averages:

$$\langle\langle Q \rangle\rangle \equiv \langle Q \rangle_0 = \frac{1}{Z_0} \int D\eta\, Q\{\eta\}\, e^{-\beta \mathcal{H}_0\{\eta\}}, \quad (13a)$$

$$\langle\langle QR \rangle\rangle = \langle QR \rangle_0 - \langle Q \rangle_0 \langle R \rangle_0 \quad (13b)$$

$$\langle\langle QRP \rangle\rangle \equiv \langle QRP \rangle_0 - \langle QR \rangle_0 \langle P \rangle_0 - \langle QP \rangle_0 \langle R \rangle_0 - \langle RP \rangle_0 \langle Q \rangle_0 + 2\langle Q \rangle_0 \langle R \rangle_0 \langle P \rangle_0 \quad (13c)$$

etc., where $Q\{\eta\}$, $R\{\eta\}$, and $P\{\eta\}$ are monomials of the field $\eta(\mathbf{x})$. Notice that the Gaussian average of any odd monomial of $\eta(\mathbf{x})$ vanishes: $\langle \eta^{2n-1}(\mathbf{x}) \rangle_0 = 0, n = 1,2,\ldots$ because the Gaussian Hamiltonian $\mathcal{H}_0\{\eta\}$ is invariant under the transformation $\eta \rightarrow (-\eta)$. Hence, $\langle \eta_V \rangle_0 = 0$. Thus, the ensemble averaged value of the volume-averaged OP is a completely anharmonic effect due to mode interactions:

$$\langle \eta_V \rangle \approx -\beta \langle\langle \eta_V \mathcal{H}_{int} \rangle\rangle + \frac{\beta^2}{2} \langle\langle \eta_V \mathcal{H}_{int}^2 \rangle\rangle = -\beta \langle\langle \eta_V \mathcal{H}_3 \rangle\rangle + \beta^2 \langle\langle \eta_V\, \mathcal{H}_3 \mathcal{H}_4 \rangle\rangle$$

$$(14)$$

Notice that the leading contribution into the 'thermal expansion' of the total order is entirely due to the cubic anharmonicity of the Hamiltonian, which is similar to the thermal expansion of solids.

The variance of the volume-averaged OP is given by:

$$Var(\eta_V) \approx \langle\langle\eta_V^2\rangle\rangle - \beta\langle\langle\eta_V^2 \mathcal{H}_{int}\rangle\rangle + \frac{\beta^2}{2}\left(\langle\langle\eta_V^2 \mathcal{H}_{int}^2\rangle\rangle - 2\langle\langle\eta_V^2 \mathcal{H}_3\rangle\rangle^2\right) \quad (15)$$

Furthermore, we present the fluctuation partition function of the system as a double product $\mathcal{Z}=\mathcal{Z}_0\mathcal{Z}_{int}$, where

$$\mathcal{Z}_0(V,\beta,\Delta) = \int D\eta \, e^{-\beta\mathcal{H}_0\{\eta\}} \quad (16a)$$

is the Gaussian partition function, and

$$\mathcal{Z}_{int}(V,\beta,\Delta) = \langle e^{-\beta\mathcal{H}_{int}\{\eta\}}\rangle_0. \quad (16b)$$

is the fluctuation-interactions partition function. For the free energy we obtain:

$$\mathcal{F} = \mathcal{F}_0 + \mathcal{F}_{int}, \quad (17)$$

where

$$\mathcal{F}_0 = -\beta^{-1}\ln \mathcal{Z}_0, \quad (18)$$

is the Gaussian approximation of the fluctuation free energy, and

$$\mathcal{F}_{int} = -\beta^{-1}\ln\langle e^{-\beta\mathcal{H}_{int}\{\eta\}}\rangle_0 \approx \langle\langle\mathcal{H}_{int}\rangle\rangle - \frac{\beta}{2}\langle\langle\mathcal{H}_{int}^2\rangle\rangle \quad (19)$$

is the fluctuation-interaction contribution up to the second order of the perturbation expansion. As expected, the first-order fluctuation-interaction correction to the free energy is the averaged interaction Hamiltonian while the second-order fluctuation-interaction correction is always negative [1]. The perturbation theory formula, Eq.(12), represents an asymptotically convergent power series, which can be differentiated termwise. Hence, the internal energy of fluctuations $\mathcal{E}=\mathcal{E}_0+\mathcal{E}_{int}$ and its variance $Var(\mathcal{E})=Var(\mathcal{E}_0)+Var(\mathcal{E}_{int})$ can be obtained from Eqs.(16-18) with the help of the thermodynamic differentiation of Eqs.(10, 11).

3. Gaussian Approximation

It is more convenient to calculate the averages in the reciprocal space of the wave vectors **k**. So, we apply the Fourier transform to the order parameter field:

$$\eta(\mathbf{x}) = \sum_{\{\mathbf{k}\}} \hat{\eta}(\mathbf{k}) e^{i\mathbf{k}\cdot\mathbf{x}} \tag{20a}$$

$$\hat{\eta}(\mathbf{k}) = \frac{1}{V} \int_V d^3x\, \eta(\mathbf{x})\, e^{-i\mathbf{k}\cdot\mathbf{x}} \tag{20b}$$

and all quantities of interest. Notice that in the reciprocal space $\eta_V = \hat{\eta}(\mathbf{0})$ and $Var(\eta_V)$ is the **k=0** component of the structure factor of the OP.

As we are interested in the bulk properties of a macroscopic cube of volume V=L³, we may adopt periodic boundary conditions for the field η(**x**). Then, independent field amplitudes η̂(**k**) have the wave vectors, which belong to the first Brillouin zone of the system: {**k**=($k_i$; i=1,2,3); $k_i$ =2π$n_i$/L; −L/2a<$n_i$≤L/2a} where $n_i$ is an integer and *a*>0 is the principle half-period of the Fourier modes that is, the *smallest relevant length scale* of the system [3-5, 18]. In crystals, $a$ is the lattice constant; in fluids, $a$ is the average inter-particle distance. Thus, the total number of the independent **k**-modes in the volume V is:

$$N = \frac{V}{a^3} \tag{21}$$

In the Gaussian approximation the **k**-modes of η̂(**k**) are independent, but they interact in the approximation described by the Hamiltonian of Eqs.(4). The latter can be expressed in terms of the field amplitudes η̂(**k**) as follows:

$$\mathcal{H}_0\{\hat{\eta}(\mathbf{k})\} = \frac{1}{2\beta} \sum_{\{\mathbf{k}\}} G^{-1}(\mathrm{k}) |\hat{\eta}(\mathbf{k})|^2 \tag{22a}$$

$$\mathcal{H}_3\{\hat{\eta}(\mathbf{k})\} = \vartheta \sum_{\{\mathbf{k}_1,\mathbf{k}_2\}} \hat{\eta}(\mathbf{k}_1)\hat{\eta}(\mathbf{k}_2)\hat{\eta}(-\mathbf{k}_1-\mathbf{k}_2) \tag{22b}$$

$$\mathcal{H}_4\{\hat{\eta}(\mathbf{k})\} = u \sum_{\{\mathbf{k}_1,\mathbf{k}_2,\mathbf{k}_3\}} \hat{\eta}(\mathbf{k}_1)\hat{\eta}(\mathbf{k}_2)\hat{\eta}(\mathbf{k}_3)\hat{\eta}(-\mathbf{k}_1-\mathbf{k}_2-\mathbf{k}_3) \tag{22c}$$

where k=|**k**| and the interaction parameters are:

$$\vartheta \equiv \frac{\Delta-3}{3} WV, \tag{23a}$$

$$u \equiv \frac{1}{2} WV. \tag{23b}$$

In Eq. (22a) we introduced the Gaussian correlation function G(k), which, using Eq.(6a) for $Q = |\hat{\eta}(\mathbf{k})|^2$, can be expressed as follows:

$$G(k) \equiv \langle |\hat{\eta}(\mathbf{k})|^2 \rangle_0 = \frac{R_0^3 \varepsilon}{V(\delta + R_0^2 k^2)} \tag{24}$$

where

$$\delta \equiv 1 - \Delta > 0, \tag{25a}$$

$$\varepsilon \equiv (\beta W R_0^3)^{-1} = k_B T / \kappa R_0, \tag{25b}$$

and

$$R_0 \equiv (\kappa/W)^{1/2} \tag{25c}$$

is the *GL radius of interactions*. The latter must be comparable to $a$— the smallest physical length scale of the system. It is reasonable to set $a=R_0$. However, for clarity, in the formulae below we leave the distinction between $R_0$ and $a$ in place.

The functional integration in Eqs.(13a, 16a) should be performed over the independent complex amplitudes of $\hat{\eta}(\mathbf{k})$ of Eq.(20a). As the field $\eta(\mathbf{x})$ is real, we have $\hat{\eta}^*(\mathbf{k}) = \hat{\eta}(-\mathbf{k})$. Hence, the mode $\hat{\eta}(\mathbf{0})$ is real and the modes $\hat{\eta}(\mathbf{k})$ with $\mathbf{k} \neq \mathbf{0}$ can be divided into two alternative sets of independent modes: those that are in the upper half-space (**k>0**) and those that are in the lower half-space (**k<0**). For definiteness, we choose the former. Furthermore, it is convenient to transform the complex modes $\hat{\eta}(\mathbf{k})$ with **k>0** to the real modes $a(\mathbf{k})$ and $b(\mathbf{k})$ as follows:

$$\hat{\eta}(\mathbf{k}) = \frac{1}{\sqrt{2}}[a(\mathbf{k}) + ib(\mathbf{k})], \tag{26a}$$

$$\hat{\eta}^*(\mathbf{k}) = \frac{1}{\sqrt{2}}[a(\mathbf{k}) - ib(\mathbf{k})]. \tag{26b}$$

and select the following set of modes: {$\hat{\eta}(\mathbf{0})$, $a(\mathbf{k})$, $b(\mathbf{k})$: **k>0**} as independent. Then the functional integration is expressed as follows:

$$\int D\hat{\eta}\,(\cdot) = \int_{-\infty}^{\infty} d\hat{\eta}(\mathbf{0}) \prod_{j=1}^{3} \prod_{\substack{k_j \le \frac{\pi}{a} \\ k_j > 0}} \int_{-\infty}^{\infty} da(\mathbf{k}) db(\mathbf{k})\,(\cdot) \qquad (27)$$

where (·) denotes an integrand. Using Eq.(27) for Eqs.(16a, 18), we obtain expressions of the partition function and free energy in the Gaussian approximation [2-5, 18]:

$$Z_0 = \prod_{\{\mathbf{k}\}} [2\pi G(k)]^{1/2} \qquad (28)$$

$$\mathcal{F}_0 = -\frac{1}{2\beta} \sum_{\{\mathbf{k}\}} \ln[2\pi G(k)] \qquad (29)$$

Using Eqs.(10, 11) for Eq.(29) we obtain expressions for the internal energy and its variance in the Gaussian approximation [3-5, 8]:

$$\mathcal{E}_0 = \frac{N}{2\beta}. \qquad (30)$$

$$Var(\mathcal{E}_0) = \frac{N}{2\beta^2} \qquad (31)$$

Eqs.(30, 31) show that, in accordance with the equipartition theorem, the Gaussian approximation of GL internal energy $\mathcal{E}_0$ is proportional to the total number of independent **k**-modes and the relative fluctuation, $Var^{1/2}(\mathcal{E}_0)/\mathcal{E}_0$, is inversely proportional to the square root of this number that is, the volume.

In large systems with L»a, the summation in Eq.(29) over the independent wave-vector modes {**k**} of the first Brillouin zone may be replaced with the integration over the continuum of the wave-vector space using the rule $\sum_{\{\mathbf{k}\}} \to \frac{V}{(2\pi)^3} \int d^3k$ . Integration over a sphere of finite radius {k≤Λ} retains the same number of independent **k**-modes as in Eq.(21) if the wave-vector cutoff is defined as $\Lambda = (6\pi^2)^{1/3}/a$. Using this cutoff and replacing summation with integration (see Eq.(A1)) we obtain:

$$\mathcal{F}_0 = \mathcal{E}_0 \left[ \ln \frac{V\delta(1+U^2)}{2\pi R_0^3 \varepsilon} - \frac{2}{3} + \frac{2}{U^2} a_1 \right], \qquad (32)$$

where

$$a_1 = 1 - \frac{\tan^{-1} U}{U}, \tag{32a}$$

$$U \equiv \Lambda R_C = \frac{(6\pi^2)^{1/3}}{\delta^{1/2}} \frac{R_0}{a}, \tag{32b}$$

is the dimensionless cutoff and

$$R_C \equiv R_0/\delta^{1/2} \tag{32c}$$

is the GL correlation length of the fluctuations.

4. Interaction Approximation

As we discussed in the Introduction, to examine metastable states only small fluctuations should be applied. This allows us to use the perturbation theory [2-5, 18] for the calculation of the interaction corrections in Eqs.(14, 19). Bearing in mind that the Gaussian averages of the odd products of the field amplitudes η̂ vanish and neglecting the terms of order $O[(\beta \mathcal{H}_{int})^3]$, we can write and expression for the volume-averaged order parameter (see Appendixes A and B):

$$\langle \eta_V \rangle \approx -3\beta \vartheta G(0) A_1 + 12\beta^2 u \vartheta G(0)[G(0) A_1^2 + 3 A_1 A_2 + 2 C_3] \tag{33}$$

In order to proceed with the analysis of Eq.(33) we have to identify small and large parameters of the perturbation expansions, Eqs.(14, 15, 19), and reveal the leading dependences of the terms on these parameters. First, one can see that δ, Eq.(25a), is not a small parameter of the expansion because we are not interested in the region near the spinodal point. Second, the traditional condition of large volume (the thermodynamic limit) is expressed here as V»$R_0^3$. Notice that the interaction parameters $u$ and $v$, Eq.(23), are not small; in fact, they are large (~V). Third, the small parameter of the perturbation expansions is ε, Eq.(25b). This means that we develop a low-temperature expansion where the average thermal energy of an independent degree of freedom is much smaller than the characteristic energy scale of the GL interactions: $k_B T$«$WR_0^3$=$\kappa R_0$. Now we can estimate the degrees of smallness of the quantities in the perturbation expansion series: β~$\varepsilon^{-1}$; G~ε/V; $A_1$~ε; $A_2$~$\varepsilon^2$/V; and $C_3$~$\varepsilon^3$/V. This leads us to the conclusion that the contributions from the second-order perturbation terms in Eq.(33) are smaller than the first-order contribution terms and our perturbation expansion is well defined. Another important fact of the perturbation expansion, Eq.(33), is that the third order perturbation expansion term $\beta^3 \langle\langle \eta_V \mathcal{H}_{int}^3 \rangle\rangle$ (more specifically $\beta^3 \langle\langle \eta_V \mathcal{H}_3^3 \rangle\rangle$) also generates the terms of the order $\varepsilon^2$. Hence, in the lowest order of expansion we have:

$$\langle \eta_V \rangle = -3\beta\vartheta G(0) A_1 + O(\varepsilon^2) = \frac{1}{\pi}\left(\frac{3}{4\pi}\right)^{1/3} \frac{2+\delta}{\delta} a_1 \frac{R_0}{a} \varepsilon + O(\varepsilon^2) \qquad (33a)$$

In the same order of approximation, the variance of the volume-averaged order parameter is expressed as follows:

$$Var(\eta_V) \approx G(0)\{1 - 12\beta u G(0) A_1 + 18\beta^2\vartheta^2 G(0)[A_2 + G(0) A_1]\}$$

$$\approx \frac{R_0^3}{V\delta}\varepsilon\left\{1 + \varepsilon\left[\frac{2}{\pi}\left(\frac{3}{4\pi}\right)^{\frac{1}{3}}\frac{R_0}{a}\left(\left(\frac{2+\delta}{\delta}\right)^2\frac{R_0}{a} - \frac{3}{\delta}\right)a_1 + \frac{(2+\delta)^2}{2\pi^2\delta^{3/2}}a_2\right]\right\} \qquad (34)$$

$$a_2 = \tan^{-1} U - \frac{U}{1+U^2} \qquad (34a)$$

As expected, $\langle \eta_V \rangle > 0$, Var($\eta_V$)~1/V, and both quantities diverge at the spinodal point.

In the same order of approximation, the fluctuation-interaction contribution to the free energy is expressed as follows:

$$\mathcal{F}_{int} \approx 3uA_1^2 - \frac{9}{2}\beta\vartheta^2 G(0) A_1^2 - 3\beta\vartheta^2 C_3 - 36\beta u^2 A_1^2 A_2 - 12\beta u^2 C_4 \qquad (35)$$

Estimating the degrees of smallness of the terms of the perturbation expansion series we obtain:

$$\mathcal{F}_{int} = -\alpha\varepsilon\mathcal{E}_0 + O(\varepsilon^3) < 0 \qquad (35a)$$

$$\alpha = \left(\frac{3}{4\pi}\right)^{\frac{2}{3}} \frac{\delta^2+\delta+4}{\pi^2\delta} \frac{a}{R_0} a_1^2 + \frac{2}{3}(2+\delta)^2 \left(\frac{a}{R_0}\right)^3 c_3 > 0 \qquad (35b)$$

$$c_3 = \int \frac{d^3x d^3y}{(2\pi)^6} \frac{1}{(1+x^2)(1+y^2)(1+|\mathbf{x}+\mathbf{y}|^2)} > 0 \qquad (35c)$$

Differentiating Eq.(35a) according to Eq.(10) we obtain that, in the leading order of expansion, the perturbative fluctuation-interaction contribution to the internal energy is $\mathcal{E}_{int} \approx -\mathcal{F}_{int}$. Hence,

$$\mathcal{E}_{int} = \alpha\varepsilon\mathcal{E}_0 + O(\varepsilon^3) \tag{36a}$$

and

$$\mathcal{E} \approx \mathcal{E}_0(1 + \alpha\varepsilon) \tag{36b}$$

Notice that the independent fluctuations and their interactions always increase the internal energy, which may not be the case for the free energy. For the variance we obtain:

$$Var(\mathcal{E}) \approx \frac{\mathcal{E}_0}{\beta}(1 + 2\alpha\varepsilon) \tag{37}$$

### III. NUMERICAL SIMULATION OF DYNAMICAL SYSTEM

#### 1. Theoretical Justification

Evolution of a fluctuating system close to the state of its equilibrium may be described by the stochastic time-dependent Ginzburg-Landau equation (STDGLE):

$$\frac{d\eta}{dt} = -\gamma\frac{\delta\mathcal{H}}{\delta\eta} + \xi(\mathbf{x}, t); \tag{38a}$$

$$\frac{\delta\mathcal{H}}{\delta\eta} \equiv \frac{\partial H}{\partial\eta} - \kappa\left(\frac{\partial^2\eta}{\partial x^2} + \frac{\partial^2\eta}{\partial y^2} + \frac{\partial^2\eta}{\partial z^2}\right) \tag{38b}$$

Here $\gamma$ is the GL relaxation coefficient and $\xi$ is the Langevin random force, which mimics the internal noise. If the noise is thermal and 'white', it obeys the following correlation conditions:

$$\langle\xi(\mathbf{x}, t)\rangle = 0, \tag{39a}$$

$$\langle\xi(\mathbf{x}, t)\xi(\mathbf{x}', t')\rangle = 2\gamma\beta^{-1}\delta(\mathbf{x} - \mathbf{x}')\delta(t - t'). \tag{39b}$$

where the averaging is over the time sequence. As known [1-4, 6], STDGLE with Eqs.(39) is consistent with the thermal equilibrium in the canonical ensemble.

If the initial condition is close to—but not exactly at—the state of equilibrium (stable or metastable), establishing equilibrium is preceded by the processes of *relaxation* whose rates depend on the length scales—*local equilibria*. Indeed, by linearizing Eq.(38) and transforming it into the Fourier space we obtain:

$$\frac{\partial \hat{\eta}(\mathbf{k},t)}{\partial t} = -\frac{1}{\tau_{rel}(|\mathbf{k}|)}\hat{\eta}(\mathbf{k},\ t) + \hat{\xi}(\mathbf{k},t); \quad (40)$$

$$\tau_{rel}(\mathbf{k}) \equiv \gamma^{-1}\beta V G(\mathbf{k}). \quad (40a)$$

where $\langle\hat{\xi}(\mathbf{k},t)\rangle = 0$ (see Eq.(39a)) and $\tau_{rel}(k)$ is the characteristic time of relaxation on the length scale $\pi/k$. The volume-averaged OP is the (**k=0**)-Fourier component of the field $\eta(\mathbf{x})$, cf. Eqs.(7, 20b). Hence, the characteristic time of establishing equilibrium in the entire system—*global equilibrium*—is given by

$$\tau_{rel} = \gamma^{-1}\beta V G(0) = \frac{1}{\gamma W \delta} \quad (40b)$$

For the purpose of comparison with the theoretical calculations, the fluctuating quantities have to be averaged. Numerically, it is not straightforward to implement the ensemble averaging according to Eq.(5). That is why the fluctuating quantities were averaged over a large observation period P»$\tau_{rel}$, which is still much smaller than the lifetime of the metastable state [7]:

$$\bar{Q} \equiv \lim_{P\to\infty}\frac{1}{P}\int_0^P dt\, Q(t)$$

$$(41)$$

Validity of the equality $\langle Q \rangle = \bar{Q}$ is justified by the Boltzmann *ergodic theorem* [17]. The time-averaged volume-averaged OP ($\bar{\eta_V}$) and the internal energy—the time-averaged effective Hamiltonian ($\mathcal{E} = \bar{\mathcal{H}}$) -- were computed together with their variances. The fluctuation free energy $\mathcal{F}$ can be calculated from $\mathcal{E}$ by the thermodynamic integration reciprocal to the differentiation of Eq.(10). For the sake of accuracy the variances were calculated as following:

$$Var(Q) = \overline{(Q-\bar{Q})^2} = \lim_{P\to\infty}\frac{1}{P}\int_0^P dt\, (Q-\bar{Q})^2 \quad (41a)$$

To compute the relaxation time of the process we used the observation period that was P«$\tau_{rel}$ that is, the system had enough time to equilibrate in small volumes (locally) but did not have enough time to equilibrate in the entire volume (globally).

2. Numerical Method

To solve Eqs.(38, 39) numerically we divide the 3D space into small cubic cells of the size $\Delta x$ on the side with **x**=($i\Delta x$, $j\Delta x$, $k\Delta x$) and time—into small steps of $\Delta t$ with t=m$\Delta t$. Then we introduce the discretized order parameter [15, 16]:

$$\eta_{ijk}^m \equiv \frac{1}{(\Delta x)^3 \Delta t} \int_{\Delta t} \int_{\Delta^3 x} dt\, d^3x\, \eta(\mathbf{x}, t) \tag{42a}$$

and Langevin force:

$$\xi_{ijk}^m \equiv \frac{1}{(\Delta x)^3 \Delta t} \int_{\Delta t} \int_{\Delta^3 x} dt\, d^3x\, \xi(\mathbf{x}, t) \tag{42b}$$

The correlation properties of the latter are the following: $\langle \xi_{ijk}^m \rangle = 0$ and

$$\langle \xi_{ijk}^m \xi_{i'j'k'}^{m'} \rangle =$$

$$\frac{1}{((\Delta x)^3 \Delta t)^2} \int_{\Delta t} \int_{(\Delta x)^3} dt\, d^3x \int_{\Delta t} \int_{(\Delta x)^3} dt'\, d^3x'\, \langle \xi(\mathbf{x}, t)\xi(\mathbf{x}', t') \rangle =$$

$$\frac{2\gamma}{\beta (\Delta x)^3 \Delta t}\, \delta_{i-i'}\delta_{j-j'}\delta_{k-k'}\delta_{m-m'} \tag{43}$$

where $\delta_l$ is the Kronecker symbol. Hence, the discretized Langevin force $\xi_{ijk}^m$ can be modeled by the Gaussian independent random number generator $R_{ijk}^m$ of zero mean, unit variance, and amplitude equal to $\sqrt{2\gamma/\beta(\Delta x)^3 \Delta t}$. Then Eqs.(39) can be solved using the explicit algorithm:

$$\eta_{ijk}^{m+1} = \eta_{ijk}^m + \frac{\gamma\kappa\Delta t}{(\Delta x)^2}\left(\eta_{i+1,jk}^m + \eta_{i-1,jk}^m + \eta_{ij+1,k}^m + \eta_{ij-1,k}^m + \eta_{ijk+1}^m + \right.$$

$$\left. \eta_{ijk-1}^m - 6\eta_{ijk}^m\right) + 2\gamma W \Delta t\, \eta_{ijk}^m(1 - \eta_{ijk}^m)\left(\eta_{ijk}^m - \frac{\delta}{2}\right) + \sqrt{\frac{2\gamma\Delta t}{\beta(\Delta x)^3}} R_{ijk}^m \tag{44}$$

Explicit solution of the boundary-value problem in the volume V=(n$\Delta$x)³ starts with the initial value $\eta_{ijk}^0 = \eta_\alpha = 0$; given the four-dimensional array $\eta_{ijk}^m$ at each point of the discretized spatial coordinates (i, j, k) on the discretized time layer m, Eq.(44) generates $\eta_{ijk}^{m+1}$ on the time layer m+1. Then the periodic boundary conditions are applied at the boundaries (i, j, k)=0 and (i, j, k)=n. Eq.(44) presents the so-called

'brute-force' method, which is a convenient tool to simulate the process of establishing equilibrium in the GL systems.

Numerical solutions of stochastic equations have a distinct feature: as we apply the independent random number generator at each grid cell at the discrete time moments, the white noise of Eqs.(39) effectively turns into the colored one with the finite *correlation length and time*: 0<(correlation length)<$\Delta x$ and 0<(correlation time)<$\Delta t$ . Thus, the superfluous grid parameters ($\Delta x$, $\Delta t$) become physical scales, which have an effect on the results of the calculations. In order for the simulation results to be comparable to the theoretical ones, the correlation length of the noise must be equal to the smallest relevant length scale $a$ of Eq.(21). In this case the reasonable choice is $a=\Delta x/2$. The best choice of the parameters $\Delta t$ is its maximum value, which still does not violate the constraint of numerical stability of Eq.(44) [19, 20]:

$$\Delta t < \frac{(\Delta x)^2}{\gamma[6\kappa+W\delta(\Delta x)^2]} \tag{45}$$

In this work we used the grid parameters: $\Delta x=2R_0$ and $\Delta t=0.5/\gamma W$. (Dependence of the results on the grid parameters will be considered in a separate publication).

For convenience, we use the following scaling for the space, time, and energy of the system:

$$\frac{x}{R_0} \to x; \quad \gamma Wt \to t; \quad \frac{\mathcal{H}}{WR_0^3} \to \mathcal{H}. \tag{46}$$

After scaling, the only control parameters that are left in the system are $\Delta$ (or $\delta$), $\varepsilon$, and $V=8n^3$. As we discussed in the Introduction, the observational period of the numerical calculations P was limited by the lifetime of the metastable state. The upper bound of the observational period depends strongly on the noise intensity $\varepsilon$, which was kept low enough to avoid the nucleation events in all simulations. In the present study P was equal to 600 and $\varepsilon \leq 0.012$. The latter practically kept the simulations in the Gaussian region of the fluctuation interactions.

3. Numerical Results

The numerical simulations started in the metastable phase ($\eta_\alpha=0$, $0<\Delta<1$) of the system of volume V with the intensity of noise $\varepsilon$. Figure 2a is a 2D slice of a typical 3D snapshot of a fluctuating OP field of the system. In Figure 2b are plotted typical time sequences of $\eta_V(t)$ and $\mathcal{H}(t)$ of the same system. In the beginning, see Fig. 2b, a fast relaxation process took place. In Figure 3 are plotted as functions of the supersaturation $\Delta$ the numerically estimated relaxation time constants at different values of the noise $\varepsilon$ and the theoretical expression of the constant for the non-interacting (linear, Gaussian) modes, $\tau_{rel}$, Eq.(40b). As expected, the relaxation time constant is practically independent of the noise but strongly dependent on the 'distance' from the spinodal point $\delta$. This is similar to the divergent behavior of a system approaching critical point of a second-order phase transition—both are due to flatness of the Hamiltonian and, hence, anomalous growth of fluctuations. In what follows the time-averaging of $\eta_V(t)$ and $\mathcal{H}(t)$ was done only on the equilibrated parts of the sequences.

In Figures 4-6 are plotted, as functions of the system size n, noise intensity $\varepsilon$, and supersaturation $\Delta$, numerically calculated time-averaged volume-averaged OP ($\overline{\eta_V}$), fluctuation internal energy ($\mathcal{E} = \overline{\mathcal{H}}$), and their respective variances Var($\eta_V$) and Var($\mathcal{E}$), together with the respective theoretical values—Eqs.(33a, 34, 36b, 37). Overall, there is a match between the theoretical and numerical results. The numerical values of the fluctuation internal energy and volume-averaged OP match their theoretical analogues better than their variances because the latter require higher accuracy of the numerical calculations (e.g. double precision). Notice that Fig. 5b can used for the graphical thermodynamic integration to obtain the fluctuation free energy $\mathcal{F}$. Of particular interest are dependencies of these quantities on the supersaturation $\Delta$. First, as theoretically predicted by Eqs.( 33a, 34), the volume-averaged OP and its variance diverge at the approach to the spinodal point, see Fig. 6a. Second, although the numerical and theoretical calculations of the internal energy and its variance match quantitatively, see Fig. 6b, dependence of the numerical quantities on the supersaturation is significantly stronger than that of the theoretical counterparts. To analyze the source of this discrepancy, we replotted in Figure 7 the results of Fig. 6b for the relative fluctuation of the internal energy Var$^{\frac{1}{2}}(\mathcal{E})/\mathcal{E}$. The numerical values match their theoretical analogues everywhere except in the domains $\Delta \to 0$ and $\Delta \to 1$. This tells us that the discrepancies of the internal energy and its variance have a common source. We believe that the source of this discrepancy is divergence of the relaxation time, Eq.(40b), correlation length, Eq.(32c), and the total order, Eq.(33a), in the vicinity of the spinodal point.

IV.     SUMMARY AND DISCUSSION

In the present publication we studied equilibrium properties of thermal fluctuations in a system governed by an asymmetric (third-power monomial in the power-series expansion), athermal (temperature independent) effective Hamiltonian in the canonical ensemble. The thermodynamic and fluctuation properties of the asymmetric Hamiltonian are different from those of the symmetric one. The most important property of the former, which is absent in the latter, is phase coexistence, which causes nucleation, growth, and coarsening phenomena. Another property of the former, which is absent in the latter, is 'thermal expansion'. The athermal Hamiltonian has two independent drivers of a phase transition—supersaturation and thermal noise, which couple in the thermal Hamiltonian.

Theoretically, we provided the low-temperature (small $\varepsilon$) perturbation expansion of the ensemble averages of the fluctuation free and internal energies, volume-averaged order parameter, and their variances. Numerically, we simulated the equilibrium fluctuations process by using the 'brute force' method and calculated the same quantities. The choice of quantities was dictated by their significance for practical purposes and conveniences for the numerical calculations. The study focused on the domain of metastability of the disordered phase of the system; however, by varying the supersaturation, the results can be extended on the domain of its absolute stability. The domain of critical behavior that is, near the spinodal point of the Hamiltonian density, was intentionally left out of the scope of the present study.

Overall, the numerical calculations match the theory within the accuracies of the methods. However, one outstanding discrepancy exists. Namely, although the numerical and theoretical calculations of the internal energy and its variance match quantitatively, the numerical dependence of these quantities on the supersaturation is significantly stronger. Attempts to improve the numerical method are underway. Results of the present study can be used for calculations of the fluctuation properties of the systems and modeling of nucleation and other rare events in the framework of the Ginzburg-Landau method.

ACKNOWLEDGMENTS

One of the authors (AU) wants to thank Mr. Guijie Wang for his help in numerical calculations and acknowledge financial assistance provided by award 70NANB14H012 from NIST U.S. Department of Commerce as part of the Center of Hierarchical Material Design at Northwestern University.

APPENDIX A: Perturbation Integrals

In this Appendix we present the definitions and calculations of the perturbation quantities. Summation over wave-vectors {**k**} was replaced with integration over the continuum of the wave-vector space using the rule $\sum_{\{\mathbf{k}\}} \to \frac{V}{(2\pi)^3} \int_0^{\Lambda} d^3k$.

$$A_0 \equiv \sum_{\{\mathbf{k}\}} \ln[2\pi G(\mathrm{k})] = N \ln \frac{2\pi R_0^3 \varepsilon}{V\delta} - \sum_{\{\mathbf{k}\}} \ln(1 + R_C^2 k^2) =$$

$$N \ln \frac{2\pi R_0^3 \varepsilon}{V\delta} - \frac{V}{(2\pi)^3} \int_0^{\Lambda} d^3\mathrm{k}\, \ln(1 + R_C^2 \mathrm{k}^2) =$$

$$N \left[ \ln \frac{2\pi R_0^3 \varepsilon}{V\delta(1+U^2)} + \frac{2}{3} - \frac{2}{U^2} + 2\frac{\arctan U}{U^3} \right] \qquad (A1)$$

$$A_{m>0} \equiv \sum_{\{\mathbf{k}\}} G^m(k) = \frac{V}{(2\pi)^3} \int_0^\Lambda d^3k\, G^m(k) = \frac{V}{2\pi^2} \left(\frac{R_0^3 \varepsilon}{V\delta}\right)^m \int_0^\Lambda \frac{dk\, k^2}{(1+R_C^2 k^2)^m} =$$

$$\frac{\varepsilon^m}{2\pi^2 \delta^{m-3/2}} \left(\frac{R_0^3}{V}\right)^{m-1} \int_0^U \frac{dz\, z^2}{(1+z^2)^m} \tag{A2}$$

$$A_1 = \frac{\varepsilon \delta^{1/2}}{2\pi^2} \int_0^U \frac{dz\, z^2}{1+z^2} = \frac{\varepsilon \delta^{1/2}}{2\pi^2} (U - \tan^{-1} U) \tag{A3}$$

$$A_2 = \frac{\varepsilon^2}{2\pi^2 \delta^{1/2}} \frac{R_0^3}{V} \int_0^U \frac{dz\, z^2}{(1+z^2)^2} = \frac{\varepsilon^2}{4\pi^2 \delta^{1/2}} \frac{R_0^3}{V} \left(\tan^{-1} U - \frac{U}{1+U^2}\right) \tag{A4}$$

$$A_{m \geq 3} = \frac{\varepsilon^m}{4\pi^2 (m-1) \delta^{m-3/2}} \left(\frac{R_0^3}{V}\right)^{m-1} \left[\int_0^U \frac{dz}{(1+z^2)^{m-1}} - \frac{U}{(1+U^2)^{m-1}}\right] \tag{A5}$$

$$C_3 \equiv \frac{V^2}{(2\pi)^6} \int_0^\Lambda d^3k\, d^3p\, G(k) G(p) G(|\mathbf{k}+\mathbf{p}|) \tag{A6}$$

$$C_4 \equiv \frac{V^3}{(2\pi)^9} \int_0^\Lambda d^3k\, d^3p\, d^3q\, G(k) G(p) G(q) G(|\mathbf{k}+\mathbf{p}+\mathbf{q}|) \tag{A7}$$

APPENDIX B: Connected Averages

$$\langle\langle \mathcal{H}_{int} \rangle\rangle = \langle\langle \mathcal{H}_4 \rangle\rangle = 3u A_1^2 \tag{B1}$$

$$\langle\langle \eta_V \mathcal{H}_{int} \rangle\rangle = \langle\langle \eta_V \mathcal{H}_3 \rangle\rangle = \vartheta \sum_{\{\mathbf{k}_1 \mathbf{k}_2\}} \langle\langle \hat{\eta}(\mathbf{0}) \hat{\eta}(\mathbf{k}_1) \hat{\eta}(\mathbf{k}_2) \hat{\eta}(-\mathbf{k}_1 - \mathbf{k}_2) \rangle\rangle =$$

$$3\vartheta G(0) A_1 \tag{B2}$$

$$\langle\langle \eta_V^2 \mathcal{H}_{int} \rangle\rangle = \langle\langle \eta_V^2 \mathcal{H}_4 \rangle\rangle =$$

$$u \sum_{\{\mathbf{k}_1 \mathbf{k}_2 \mathbf{k}_3\}} \langle\langle \hat{\eta}(\mathbf{0}) \hat{\eta}(\mathbf{0}) \hat{\eta}(\mathbf{k}_1) \hat{\eta}(\mathbf{k}_2) \hat{\eta}(\mathbf{k}_3) \hat{\eta}(-\mathbf{k}_1 - \mathbf{k}_2 - \mathbf{k}_3) \rangle\rangle =$$

$$12 u G^2(0) A_1 \tag{B3}$$

$$\langle\langle \mathcal{H}_0 \mathcal{H}_{int} \rangle\rangle = \langle\langle \mathcal{H}_0 \mathcal{H}_4 \rangle\rangle =$$

$$\frac{u}{2\beta} \sum_{\{\mathbf{k}\}} \sum_{\{\mathbf{k}_1 \mathbf{k}_2 \mathbf{k}_3\}} \langle\langle \hat{\eta}(\mathbf{k})\hat{\eta}(-\mathbf{k})\hat{\eta}(\mathbf{k}_1)\hat{\eta}(\mathbf{k}_2)\hat{\eta}(\mathbf{k}_3)\hat{\eta}(-\mathbf{k}_1 - \mathbf{k}_2 - \mathbf{k}_3) \rangle\rangle =$$

$$6 \frac{u}{\beta} A_1^2 \tag{B4}$$

$$\langle\langle \mathcal{H}_{int}^2 \rangle\rangle = \langle\langle \mathcal{H}_3^2 \rangle\rangle + \langle\langle \mathcal{H}_4^2 \rangle\rangle \tag{B5}$$

$$\langle\langle \mathcal{H}_3^2 \rangle\rangle = \vartheta^2 \{9G(0)A_1^2 + 6C_3\} \tag{B6}$$

$$\langle\langle \mathcal{H}_4^2 \rangle\rangle = u^2 \{12 A_1^2 A_2 + 24 C_4\} \tag{B7}$$

$$\langle\langle \eta_V \mathcal{H}_{int}^2 \rangle\rangle = 2\langle\langle \eta_V \mathcal{H}_3 \mathcal{H}_4 \rangle\rangle =$$

$$2\vartheta u \sum_{\{\mathbf{k}_1 \mathbf{k}_2 \mathbf{k}_3 \mathbf{k}_4 \mathbf{k}_5\}} \langle\langle \hat{\eta}(\mathbf{0})\hat{\eta}(\mathbf{k}_1)\hat{\eta}(\mathbf{k}_2)\hat{\eta}(-\mathbf{k}_1 - \mathbf{k}_2)\hat{\eta}(\mathbf{k}_3)\hat{\eta}(\mathbf{k}_4)\hat{\eta}(\mathbf{k}_5)\hat{\eta}(-\mathbf{k}_3 - \mathbf{k}_4 - \mathbf{k}_5) \rangle\rangle = 24\vartheta u G(0)\{G(0) A_1^2 + 3 A_1 A_2 + 2 C_3\} \tag{B8}$$

$$\langle\langle \mathcal{H}_0^2 \mathcal{H}_{int} \rangle\rangle = \langle\langle \mathcal{H}_0 \mathcal{H}_4 \rangle\rangle = 18 \frac{u}{\beta^2} A_1^2 \tag{B9}$$

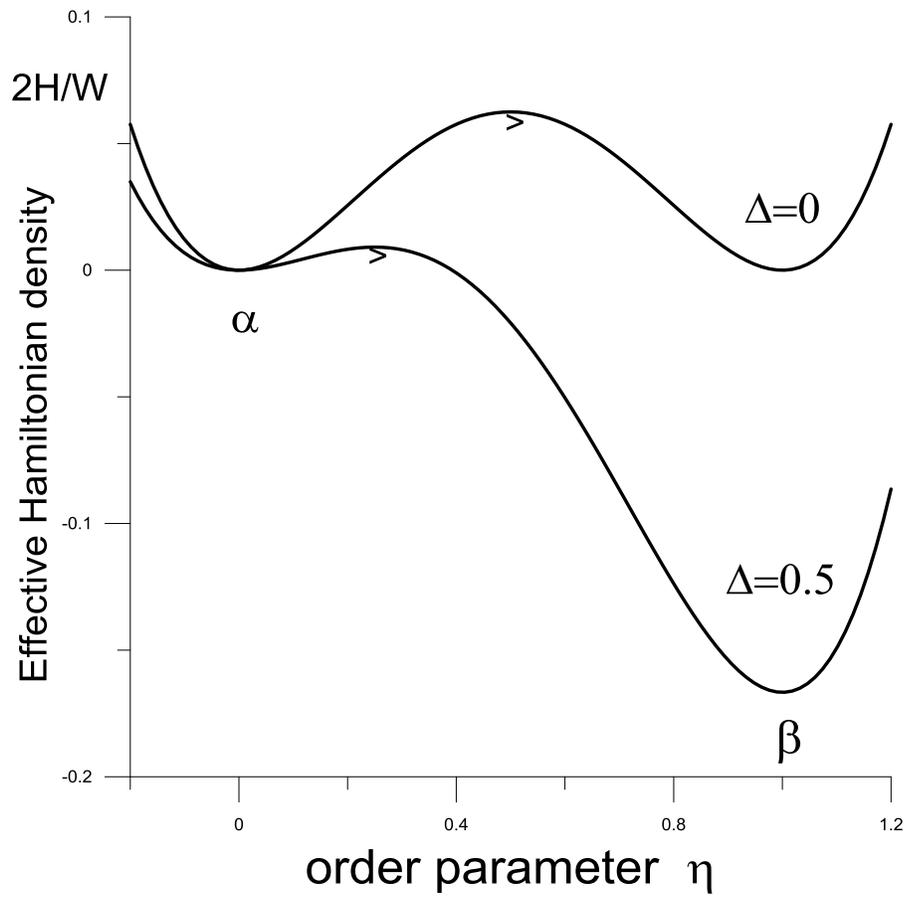

Figure 1. Effective Hamiltonian density, Eq. (2), as a function of the order parameter $\eta$ for two values of the supersaturation $\Delta$. $\alpha$ and $\beta$ are metastable and stable phases; crosses are the transition state.

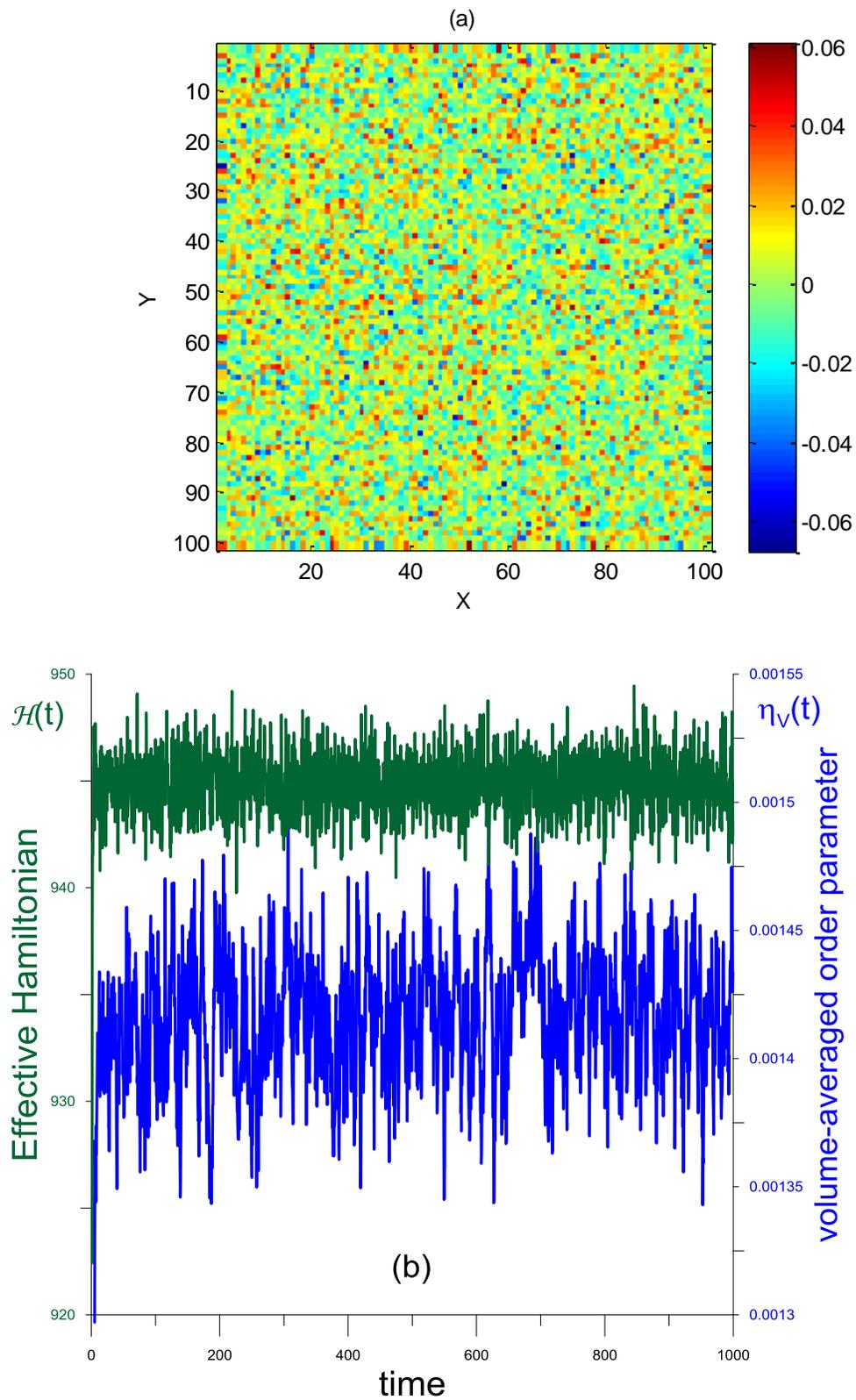

Figure 2. (Color online) (a) 2D slice of the 3D snapshot of a fluctuating OP field and (b) time sequences of $\eta_V(t)$ (blue line) and $\mathcal{H}(t)$ (green line) in the system of $V=8\times10^6$, $\varepsilon=2\times10^{-3}$, and $\Delta=0.5$.

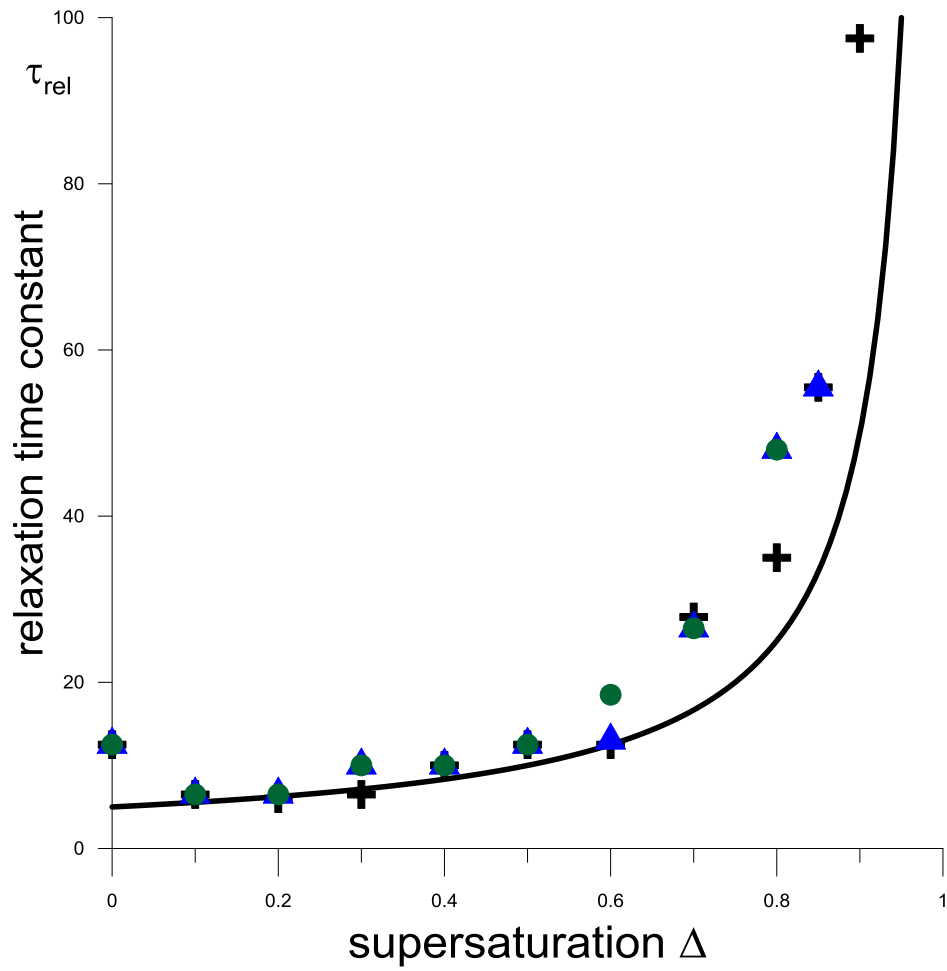

Figure 3. (Color online) Dependence of the relaxation time constant on the supersaturation $\Delta$ at $V=8\times10^6$. Numerical results: black crosses—$\varepsilon=2\times10^{-3}$, blue triangles—$\varepsilon=5\times10^{-3}$, green circles—$\varepsilon=1\times10^{-2}$. Curve—theoretical calculation, Eq.(40b).

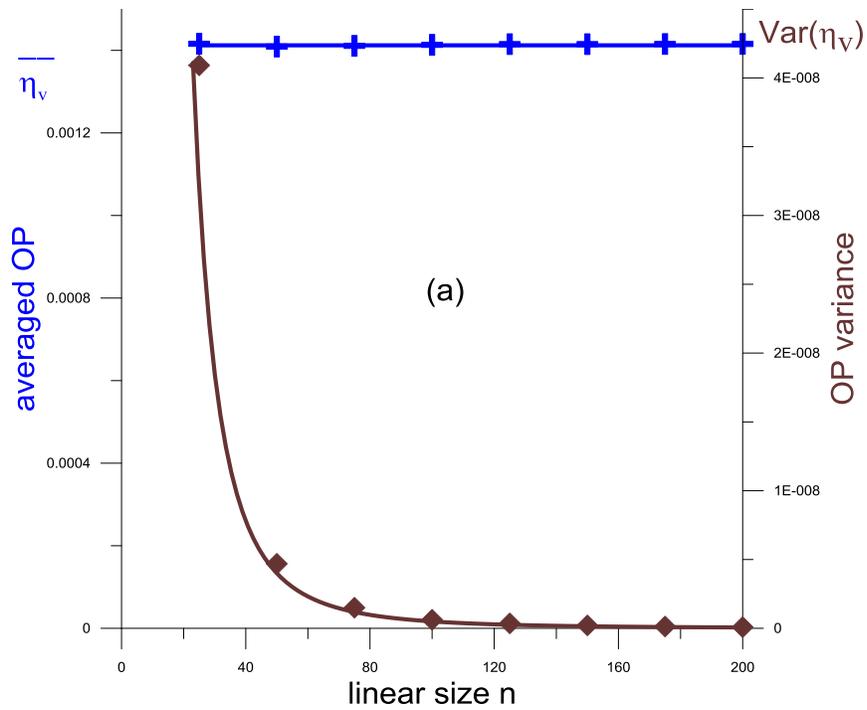

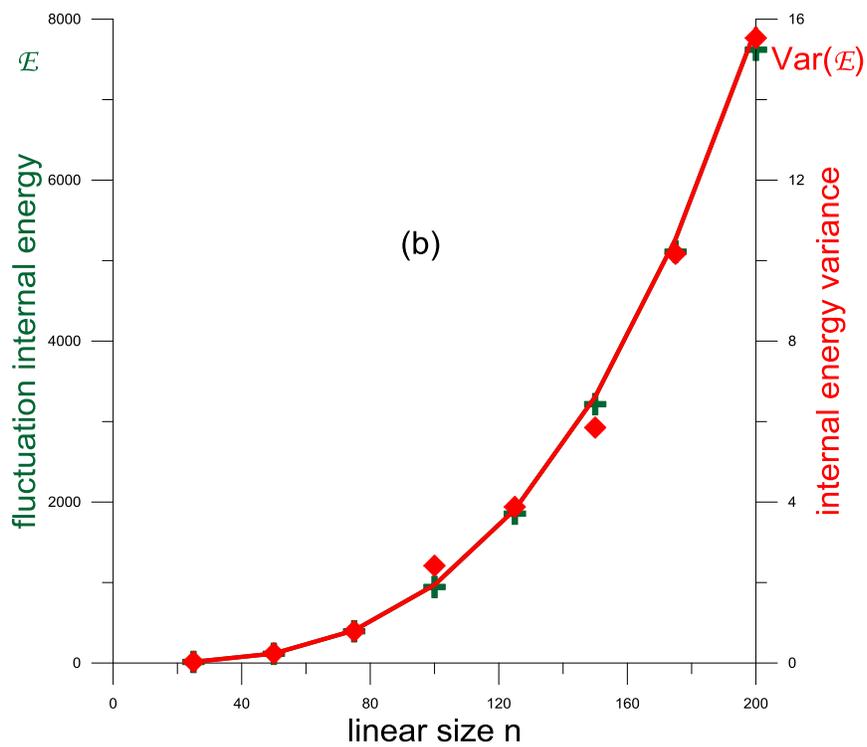

Figure 4. (Color online) Dependences of (a) averaged order parameter (blue line) and its variance (brown line) and (b) internal energy (green line) and its variance (red line) on the linear size n of the system at $\varepsilon=2\times10^{-3}$ and $\Delta=0.5$. Crosses—numerical results, curves and straight lines—theoretical calculations.

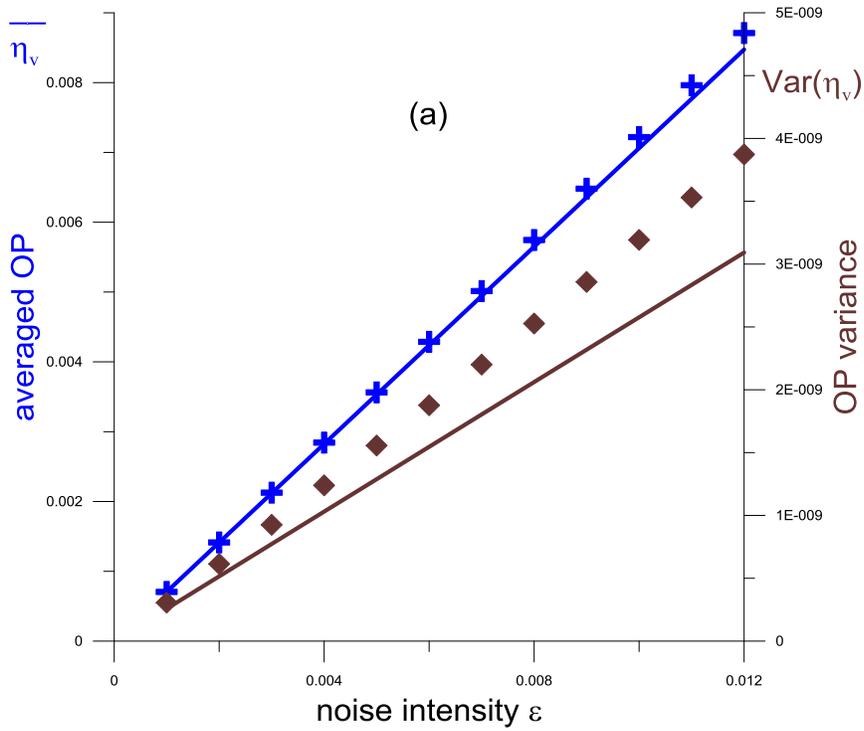

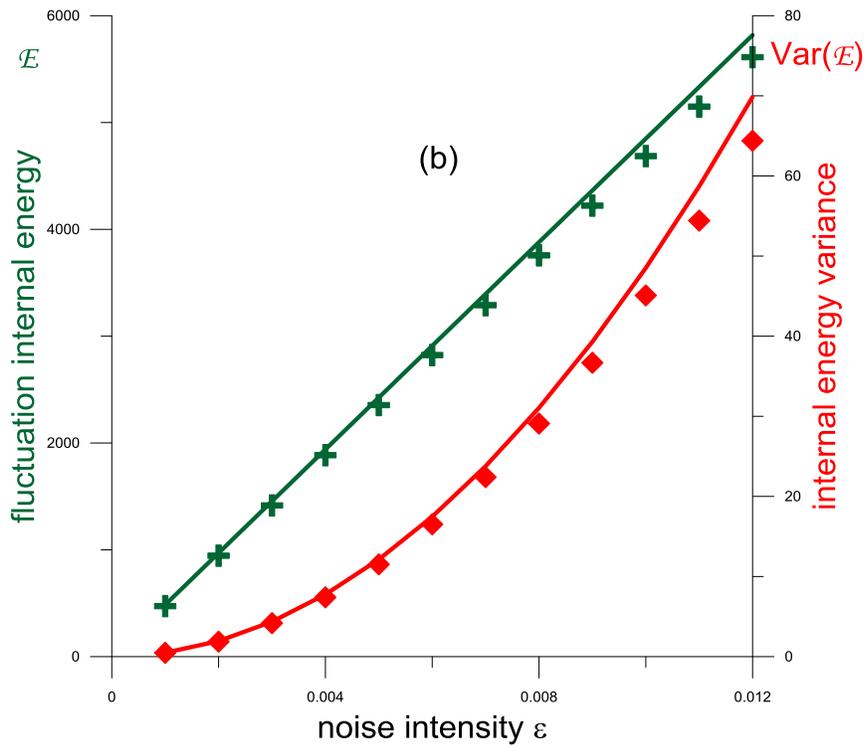

Figure 5. (Color online) Dependences of (a) averaged order parameter (blue line) and its variance (brown line) and (b) internal energy (green line) and its variance (red line) on the noise intensity $\varepsilon$ at $V=8\times10^6$ and $\Delta=0.5$. Crosses—numerical results, curves and straight lines—theoretical calculations.

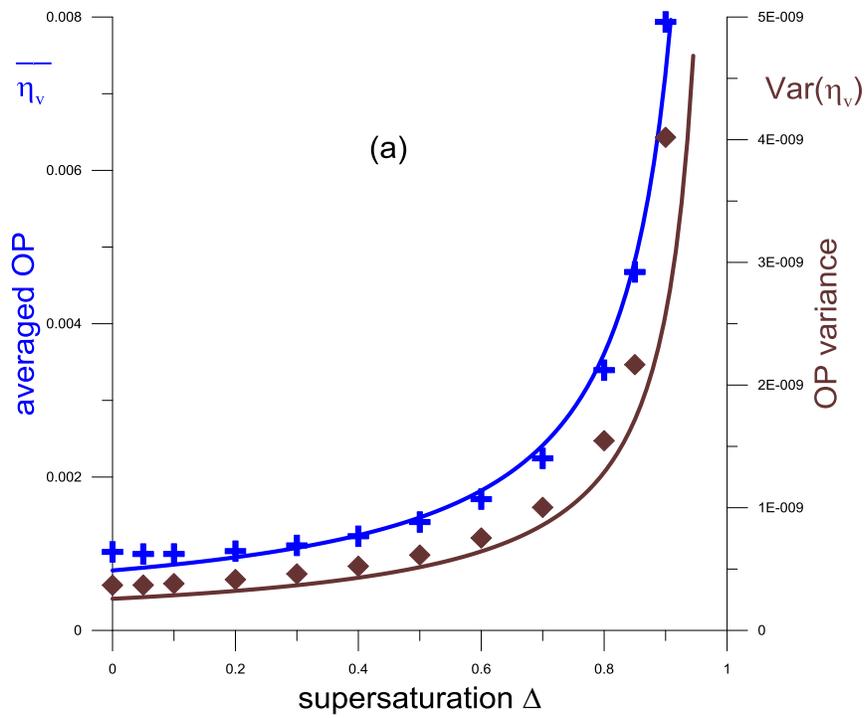

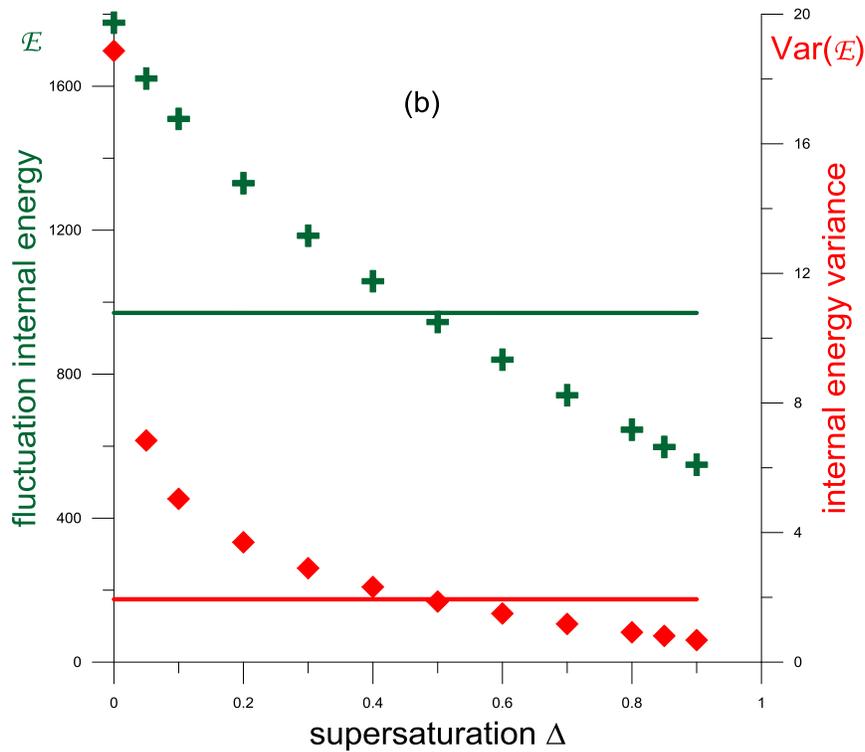

Figure 6. (Color online) Dependences of (a) averaged order parameter (blue line) and its variance (brown line) and (b) internal energy (green line) and its variance (red line) on the supersaturation $\Delta$ at $V=8\times10^6$ and $\varepsilon=2\times10^{-3}$. Crosses—numerical results, curves and straight lines—theoretical calculations.

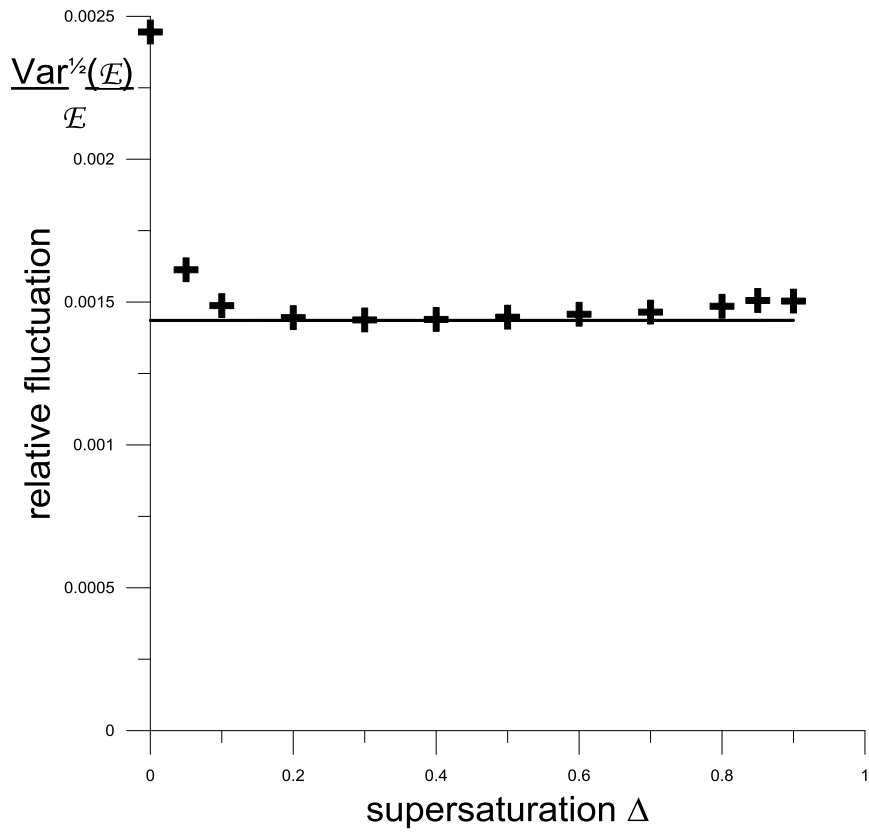

Figure 7. Relative fluctuation of the internal energy as a function of supersaturation. Crosses—numerical results, straight line—theoretical calculations.